\documentclass{PoS}
\usepackage{amsmath}

\title{Two--particle correlations and balance functions in p--Pb and Pb--Pb collisions at LHC energies with ALICE}

\ShortTitle{Two--particle correlations and balance functions in p--Pb and Pb--Pb}

\author{\speaker{Michael WEBER} on behalf of the ALICE collaboration\\%
       University of Houston\\
       E-mail: \email{m.weber@cern.ch}}
       
\abstract{Recent measurements of two--particle correlations in high--multiplicity p--Pb collisions at $\sqrt{s_{NN}} = 5.02$~TeV revealed a long--range structure (large separation in $\Delta\eta$) at the near-- ($\Delta\varphi\simeq0$) and away--side ($\Delta\varphi\simeq\pi$) of the trigger particle. At LHC energies, these ridge--like structures have not only been observed in Pb--Pb collisions, but also in high--multiplicity pp collisions. In the first case, this phenomenon is commonly related to collectivity in hadron production, i.e. hydrodynamic evolution, whereas in the latter, mechanisms like longitudinal color connections and multi--parton interactions might play an important role. To shed light on the particle production mechanisms in p--Pb and Pb--Pb collisions and answer the question about collectivity, we extend the two--particle correlation analysis for hadrons in two directions: identified particles, which should show a characteristic pattern in case of collective motion in a hydrodynamic medium, and charge--dependent correlations studied with the balance function, which are sensitive to charge--dependent effects like local charge conservation.}

\FullConference{The European Physical Society Conference on High Energy Physics \\
		 18-24 July, 2013\\
		 Stockholm, Sweden}

\begin{document}

\section{Introduction}
\noindent Strongly interacting matter at extreme conditions, like high temperatures and/or densities, is probed in the laboratory in ultrarelativistic heavy--ion collisions. Quantum chromodynamics (QCD) \cite{Ref:QCD} predicts that at sufficiently high energy density, of the order of 0.5~GeV/fm$^3$ \cite{Ref:QGP}, a deconfined state of quarks and gluons is produced, the so-called quark gluon plasma (QGP). First results from the ALICE detector \cite{Ref:Alice} at the Large Hadron Collider (LHC) at CERN were obtained from Pb--Pb collisions at $\sqrt{s_{\mathrm{NN}}}=2.76$~TeV \cite{Ref:LHCHI}. The reference measurements from p--Pb collisions at $\sqrt{s_{\mathrm{NN}}}=5.02$~TeV are needed to disentangle cold nuclear matter effects, like nuclear parton distribution functions in the initial state and rescattering on nucleus constituents in the final state, from the anticipated hot QCD matter effects. 
Two--particle correlations, i.e. measuring the distributions of relative angles of two particles in azimuthal angle $\Delta\varphi$ and pseudorapidity $\Delta\eta$,  are a powerful tool to study the underlying mechanism and the dynamics of particle production in collisions of nuclei and nucleons.  In high--multiplicity p--Pb collisions at $\sqrt{s_{NN}} = 5.02$~TeV a long--range structure (extended $\Delta\eta$) at the near-- ($\Delta\varphi\simeq0$) and away--side ($\Delta\varphi\simeq\pi$) of the trigger particle was revealed \cite{Ref:ALICEpPbCorr}. These ridge structures can attributed to mechanisms that involve initial-state effects, such as gluon saturation \cite{Ref:GluonSaturation} and color connections forming along the longitudinal direction \cite{Ref:CR}, and final state effects, such as parton--induced interactions \cite{Ref:PartonInducedInteractions}, and collective effects developing in a high--density system possibly formed in these collisions \cite{Ref:Collective}.
In order to further characterize this effect two observables are presented in this paper: the Fourier decomposition \cite{Ref:Fourier} of the long--range structure of two--particle correlations for pions, kaons and protons and the charge--dependent correlation functions, known as balance functions \cite{Ref:BF_theory}.

\section{Two--particle correlations for identified particles}

\noindent In order to study the particle type dependence of two--particle correlations, unidentified charged tracks are combined with pions, kaons or protons (denoted as $h-\pi$, $h-K$ or $h-p$ respectively). The particle identification is based on the difference between the expected and measured signal in the Time Of Flight (TOF) detector and from the specific energy loss in the Time Projection Chamber (TPC) and is expressed in multiples of the detector resolution $\sigma$ - $N_{\sigma}$. For a given species, only particles that fulfill the requirement
\begin{equation}
3>N_{\sigma}=
	\begin{cases}
     		\sqrt{N_{\sigma,TOF}^2+N_{\sigma,TPC}^2} & \text{, if $p_{\mathrm{T}}>0.5\mathrm{~GeV/}c$}.\\
    		N_{\sigma,TPC} & \text{, if $p_{\mathrm{T}}<0.5\mathrm{~GeV/}c$}.
  	\end{cases}
 \end{equation}
are selected, which yields in a low contamination ($<15\%$) from particle type misidentification.
The two--particle correlation is expressed by the associated yield per trigger particle in a transverse momentum $p_{\mathrm{T}}$ interval:
\begin{equation}
\frac{1}{N_{\mathrm{trig}}}\frac{d^2N_{\mathrm{assoc}}}{d\Delta\eta\Delta\varphi}=\frac{S(\Delta\eta,\Delta\varphi)}{B(\Delta\eta,\Delta\varphi)}
\end{equation}
where $N_{\mathrm{trig}}$ is the total number of trigger particles in the event class and $p_{\mathrm{T}}$ interval. The signal distribution $S(\Delta\eta,\Delta\varphi)=\frac{1}{N_{\mathrm{trig}}}\frac{d^2N_{\mathrm{same}}}{d\Delta\eta\Delta\varphi}$ is the associated yield for particle pairs from the same event. The background distribution $B(\Delta\eta,\Delta\varphi) = \alpha \frac{d^2N_{\mathrm{mixed}}}{d\Delta\eta\Delta\varphi}$ corrects for pair acceptance and pair efficiency and is constructed by correlating trigger particles in one event with the associated particles from other events.
Based on the charge deposition in the VZERO-A ($2.8<\eta<5.1$) detector, the analyzed events are divided into four multiplicity classes "0-20\%", "20-40\%", "40-60\%" and "60-100\%" from highest to lowest multiplicities. In order to eliminate or reduce the influence of particle correlations from parton hadronization in jets, the results from the lowest multiplicity class "60-100\%" (where "jet-like" structures dominate) are subtracted. The projection on the $\Delta\varphi$ axis and harmonic decomposition is done from a fit with:
\begin{equation}
\frac{1}{N_{\mathrm{trig}}}\frac{dN_{\mathrm{assoc}}}{\Delta\varphi}= a_0 +2a_1\cos(\Delta\varphi)+2a_2\cos(2\Delta\varphi)+2a_3\cos(3\Delta\varphi).
\end{equation}
\noindent Since a residual peak structure is observed in the remaining correlation functions, the projection is averaged over $0.8 < |\Delta\eta| < 1.6$ on the near side and $|\Delta\eta| < 1.6$ on the away side. The $v_n$ components for the subtracted two--particle correlations, i.e. $v_n\{\mathrm{2PC, sub}\}$, can be extracted from the Fourier coefficients for pions, kaons and protons. A detailed description of the analysis can be found in \cite{Ref:ALICEpPbCorr,Ref:ALICEpPbCorrPID}.

\begin{figure*}[htb]
\centering
\includegraphics[width=0.75\linewidth]{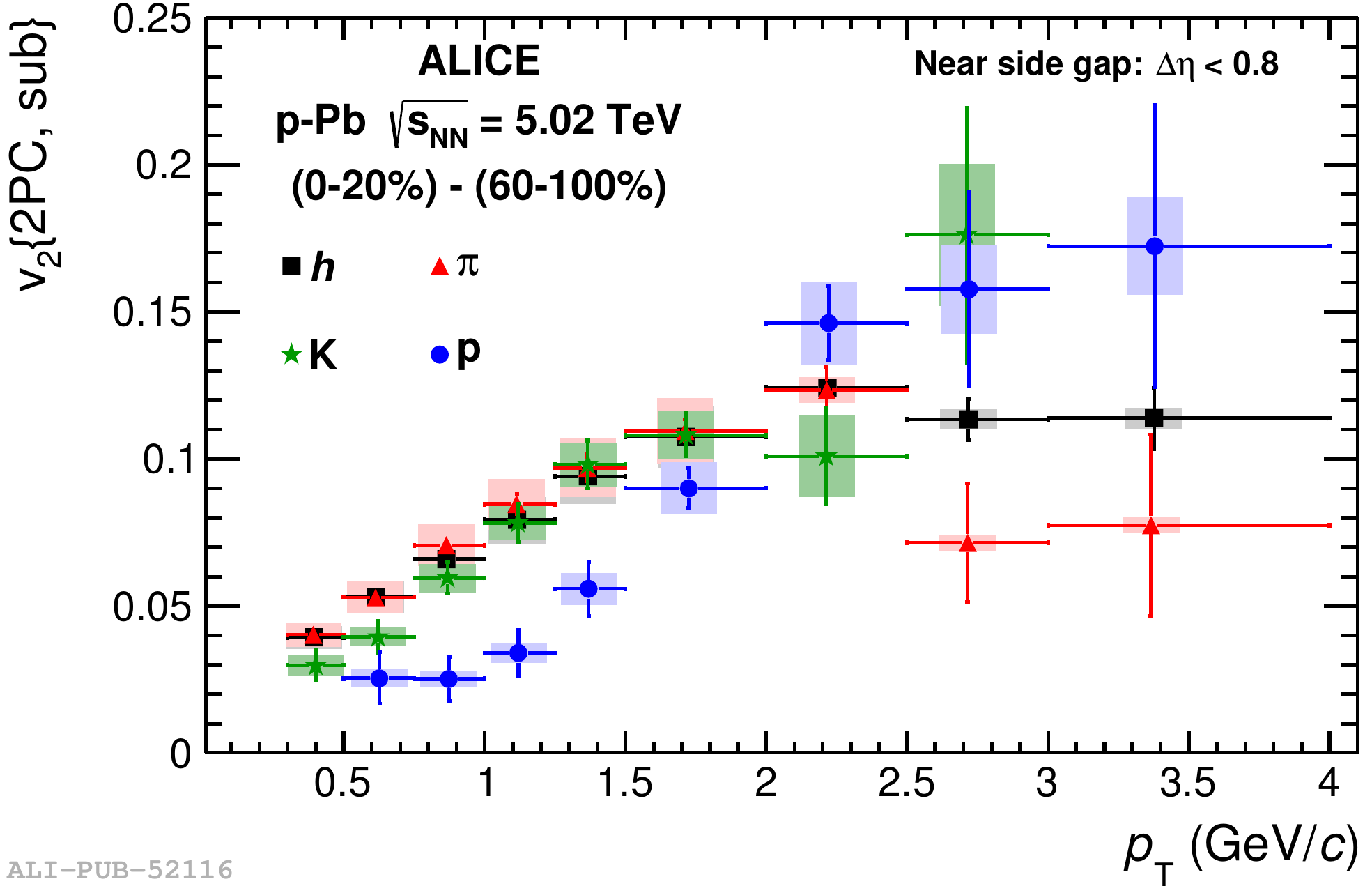}
\caption{The Fourier coefficient $v_2\{\mathrm{2PC, sub}\}$ for hadrons (black squares), pions (red triangles), kaons (green stars) and protons (blue circles) as a function of $p_{\mathrm{T}}$ from the correlation in the 0-20\% multiplicity class after subtraction of the correlation from the 60-100\% multiplicity class. Error bars show statistical uncertainties while shaded areas denote systematic uncertainties. Figure from \cite{Ref:ALICEpPbCorrPID}.}
\label{fig:CFresults}
\end{figure*}
\noindent Fig.~\ref{fig:CFresults} shows the Fourier coefficient $v_2\{\mathrm{2PC, sub}\}$ for all particle species as a function of $p_{\mathrm{T}}$. For $p_{\mathrm{T}}< 2$~GeV/$c$ a clear mass ordering $v_{2,\pi}>v_{2,p}$ is observed, for $p_{\mathrm{T}}>2$~GeV/$c$ the Fourier coefficients hint at an inversion of this mass ordering. This behavior is qualitatively similar to the findings in Pb--Pb collisions \cite{Ref:ALICEflowPbPb}.

\section{Balance functions}

In order to study the charge dependence of the two--particle correlation function one uses the balance function definition \cite{Ref:BF_theory}:
\begin{equation}
B_{+-}(\Delta \eta,\Delta \varphi)=\left(\frac{S(\Delta\eta,\Delta\varphi)}{B(\Delta\eta,\Delta\varphi)}\right)_{\mathrm{US}}-\left(\frac{S(\Delta\eta,\Delta\varphi)}{B(\Delta\eta,\Delta\varphi)}\right)_{\mathrm{LS}}
\end{equation}
\noindent where US denotes the correlations for unlike sign pairs, i.e. combinations of positive--negative charge, and LS for like sign pairs, i.e. positive--positive and negative--negative charge. Assuming that unlike sign pairs are created at the same space--time point and correlated in momentum space due to a strong collective expansion, their separation in pseudorapidity $\Delta\eta$ and $\Delta\varphi$ depends not only on the initial momentum difference, but also on the length of the rescattering phase. Furthermore, it was shown that the balance function for the relative azimuthal angle of the charge--anticharge pair can probe the collective motion of the produced system and, in particular, its radial flow \cite{Ref:BF_phi}. 
First results for the balance functions at LHC energies were obtained for unidentified charged particles in Pb--Pb collisions at $\sqrt{s_{\mathrm{NN}}}=2.76$~TeV, the details of the analysis can be found in \cite{Ref:ALICEBF}.
\begin{figure*}[htb]
\centering
\includegraphics[width=0.6\linewidth]{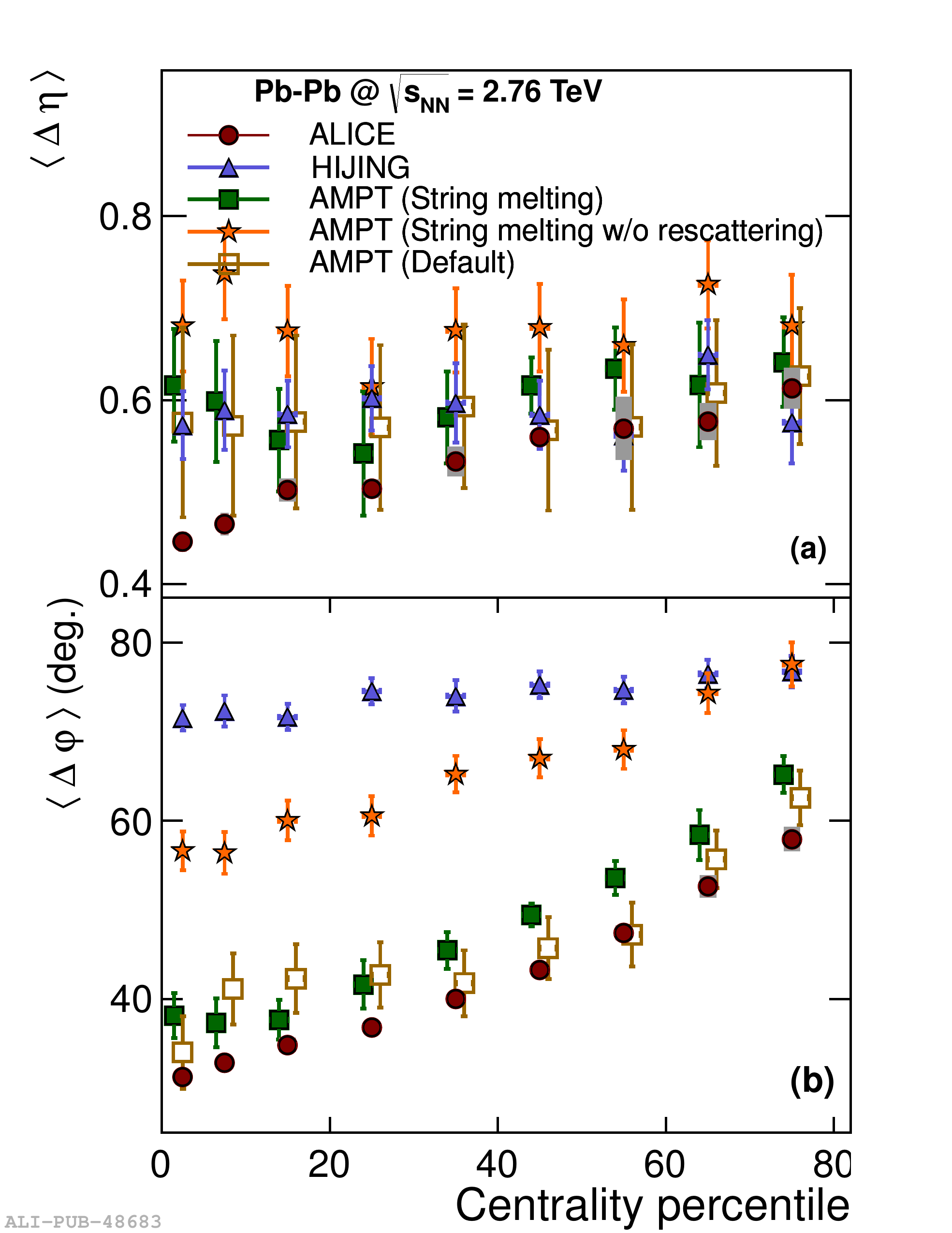}
\caption{(color online).  The centrality dependence of the width of the balance function $\langle \Delta \varphi \rangle$~
and $\langle \Delta \varphi \rangle$, for the correlations studied in terms of the relative pseudorapidity and 
the relative azimuthal angle, respectively \cite{Ref:ALICEBF}. The data points are compared to the predictions from HIJING \cite{Ref:Hijing} and AMPT \cite{Ref:Ampt}. Figure from \cite{Ref:ALICEBF}.}
\label{fig:BF_PbPb}
\end{figure*}

\noindent The weighted averages
\begin{eqnarray}
\langle \Delta \eta \rangle = \sum_{i=1}^k{[B_{+-}(\Delta \eta _i) \cdot \Delta \eta _i]}/\sum_{i=1}^k{B_{+-}(\Delta \eta _i)},   \\                                                   
\langle \Delta \varphi \rangle = \sum_{i=1}^k{[B_{+-}(\Delta \varphi _i) \cdot \Delta \varphi _i]}/\sum_{i=1}^k{B_{+-}(\Delta \varphi _i)},
\end{eqnarray}
\noindent where $B_{+-}(\Delta \eta _i)$ is the balance function value for each bin $\Delta \eta_i$, with the sum running over all bins $k$, are shown in Fig.~\ref{fig:BF_PbPb}. A clear narrowing going from peripheral to central collisions in $\Delta\eta$ and $\Delta\varphi$ is observed, which is not reproduced by different MC event generators. HIJING \cite{Ref:Hijing} produces charges early in the collision history, mainly via string fragmentation, and includes no collective motion, e.g. radial flow. AMPT (string melting) \cite{Ref:Ampt}, on the other hand, with parameters tuned to reproduce the measured elliptic flow values of non--identified particles at the LHC, shows a qualitative agreement in the centrality trend for $B(\Delta \varphi)$. This can be understood as collective flow serving as the determining source of balancing charge correlation in $\Delta\varphi$.\\
Charge--dependent correlation functions from event--by--event, 3+1D, viscous hydrodynamical calculations which include local charge conservation also show a difference between LS and US pairs in heavy--ion collisions \cite{Ref:HydroPbPb}. Furthermore, a near--side ridge in p--Pb collisions is reproduced as well \cite{Ref:HydropPb}, so a natural test of hydrodynamical models of this type is the comparison of the balance function in this collision system.
\begin{figure*}[htb]
\centering
\includegraphics[width=0.45\linewidth]{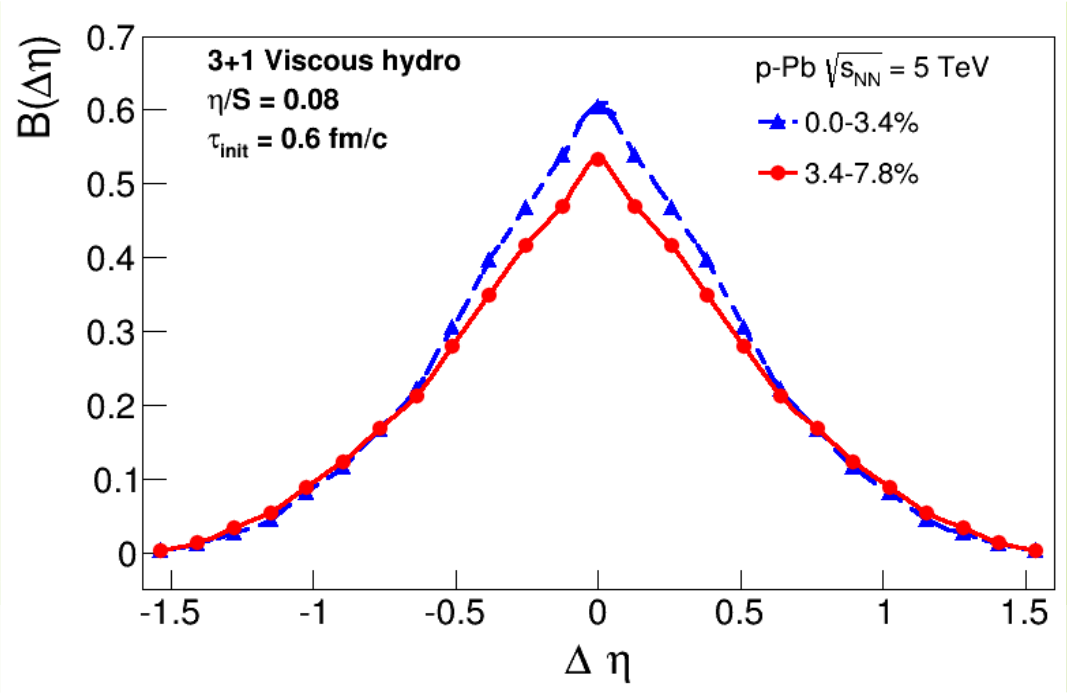}
\includegraphics[width=0.45\linewidth]{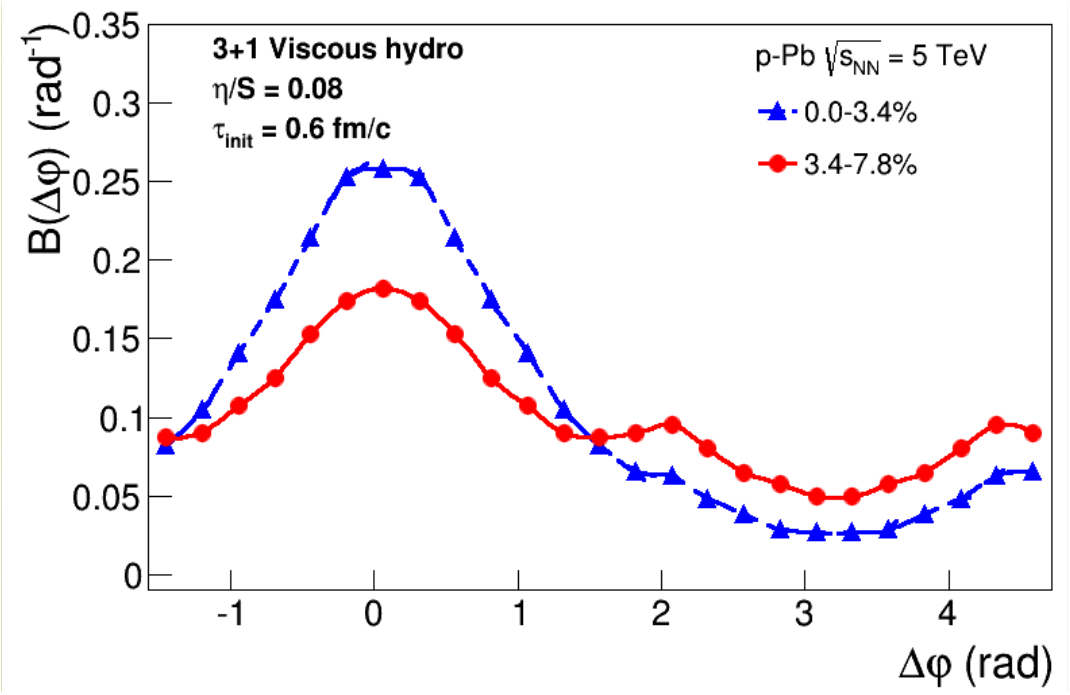}
\caption{Balance function in $\Delta\eta$ and $\Delta\varphi$ for charged particles in p--Pb collisions at $\sqrt{s_{\mathrm{NN}}}=5.02$~TeV for two different centrality intervals from  event--by--event, 3+1D, viscous hydrodynamical calculations which include local charge conservation \cite{Ref:HydroPbPb,Ref:HydropPb,Ref:HydropPbBF}.}
\label{fig:BF_pPb}
\end{figure*}

Fig.~\ref{fig:BF_pPb} shows the balance function in $\Delta\eta$ and $\Delta\varphi$ in p--Pb collisions at $\sqrt{s_{\mathrm{NN}}}=5.02$~TeV from the hydrodynamical calculations \cite{Ref:HydropPbBF}. A clear dependence on the centrality is predicted: the width in relative azimuthal angle and pseudorapidity decreases when going from peripheral to central p--Pb collisions. In addition, a sensitivity to the input parameters, like the shear viscosity over entropy density ratio ($\eta/s$) of the produced medium and the time of onset of the hydrodynamical regime ($\tau_{\mathrm{init}}$), is expected \cite{Ref:HydropPbBF}.

\section{Summary}

\noindent The origin of the long--range structure (large separation in $\Delta\eta$) at the near-- and away--side of the trigger particle in two--particle correlations in high--multiplicity p--Pb collisions at $\sqrt{s_{NN}} = 5.02$~TeV is further studied in two extensions. The Fourier coefficient $v_2$ for pions, kaons and protons, extracted from two--particle correlations, shows a clear mass ordering at low momenta $p_{\mathrm{T}}< 2$~GeV/$c$, which is qualitatively similar to the results from Pb--Pb collisions. The charge--dependent correlation functions, also known as the balance function, from event--by--event, 3+1D, viscous hydrodynamical calculations which include local charge conservation show a clear centrality dependence in their width in $\Delta\eta$ and $\Delta\varphi$ and a sensitivity to the input parameters. Both measurements offer the unique opportunity to test the hydrodynamical models for p--Pb collisions extensively.

\end{document}